# Response to "Horizontal gene transfer may explain variation in θs"

*Iñigo Martincorena, Nicholas M. Luscombe*

**Abstract**

In a short article submitted to ArXiv [1], Maddamsetti *et al*. argue that the variation in the neutral mutation rate among genes in *Escherichia coli* that we recently reported [2] might be explained by horizontal gene transfer (HGT). To support their argument they present a reanalysis of synonymous diversity in 10 *E.coli* strains together with an analysis of a collection of 1,069 synonymous mutations found in repair-deficient strains in a long-term in vitro evolution experiment. Here we respond to this communication. Briefly, we explain that HGT was carefully accounted for in our study by multiple independent phylogenetic and population genetic approaches, and we show that there is no new evidence of HGT affecting our results. We also argue that caution must be exercised when comparing mutations from repair deficient strains to data from wild-type strains, as these conditions are dominated by different mutational processes. Finally, we reanalyse Maddamsetti's collection of mutations from a long-term in vitro experiment and we report preliminary evidence of non-random variation of the mutation rate in these repair deficient strains.

**Introduction**

In their communication Maddamsetti and colleagues [1] present two main arguments:
1. *Horizontal gene transfer (HGT) dominates synonymous diversity*: They analysed synonymous diversity in 10 *E.coli* strains and report that HGT causes higher diversity.
2. *Uniform mutation rates in a long-term experiment*: They also analysed the distribution of 1,069 synonymous mutations from their long-term experiments and suggest that mutation rates are uniform along the genome.

From these observations, Maddamsetti concludes that the variation reported in our study might be explained by HGT. Below we explain why the analyses and conclusions are incorrect.

**1. Evidence of HGT is only present in Maddamsetti et al. data, not in our published estimates**

To test for evidence of HGT in synonymous diversity in *E.coli*, Maddamsetti calculated their own estimates of θs using 10 published *E.coli* genomes, without applying any of the filters that we carefully designed to avoid HGT (see Supplementary Information SI 2.2). Unsurprisingly, they found an association between θs and HGT (*P*=0.022), whereas repeating the same analysis on our published data does not yield any association (*P*>0.10).

The potential effects of HGT were carefully considered in our study, and they are extensively described in the Supplementary Information (SI 2.2, 3.1.4, 3.2, 6.1). Neither distant HGT events nor those between closely related strains affect the results presented in our study. Distant HGT events are easy to identify as they elongate the corresponding branch in the alignment tree, as well as causing increased diversity and divergence. Any genes affected by distant HGT were removed by our filters before any analysis (SI 2.2 for details). HGT among close strains was also carefully quantified (referred to as homologous recombination for consistency with the population genetic literature). We provided extensive evidence that the functional variation of θs is independent of recombination rate (SI 3.2 and SI 6.1) and any potential impact of background selection and hitchhiking (SI 3.1.4).

**2. Data from repair-deficient strains cannot be compared directly with observations of mutational processes dominating only in the wild-type**

Having performed the above analysis, Maddamsetti examined 1,069 synonymous mutations from Lenski's long-term evolutionary experiment. Unfortunately, 1,055 (>98%) of these mutations came from repair-deficient strains with exceptionally high mutation rates (around two orders of magnitude above the wildtype) and dominated by very different mutational processes (very atypical mutational spectra compared with wildtype). In such conditions, a genome could be flooded with largely randomly occurring mutations, masking any underlying non-random features present in wildtype strains. Therefore, any findings from repair-deficient strains must be interpreted with caution and cannot be used to prove or disprove the conclusions of our study about mutational processes that dominate in the wild-type.

## 3. Inconclusive analysis of uniform mutation rate

In addition, the analyses presented by Maddamsetti suffer from several basic flaws.

(i) First, the heterogeneous model should be expressed as mutations_per_gene $\propto \theta_s*$gene_length instead of mutations_per_gene $\propto \theta_s$, since $\theta_s$ is an estimate of the mutation rate per base. Maddamsetti's analysis is therefore invalid. (ii) Second, the use of $\theta_s$ to sort the x-axis is misleading as it is a noisy variable (see SI 6.1) and Maddamsetti's analysis implicitly expects the data to fit this noise. Indeed, the cumulative plots used by Maddamsetti would always appear to fit the uniform model perfectly even if there are huge variations in the mutation rate, as long as these variations are not strongly correlated with our estimates of $\theta_s$. This is easily demonstrated by simulation (Figure 1). Consequently, uniformity cannot be concluded from Maddamsetti's analysis.

In our study, we presented a robust test for uniform mutation rates using Monte Carlo simulations (SI 6.1). We repeated this analysis on Maddamsetti's dataset; unfortunately owing to the limited number of mutations and the dominating effect of gene length (see above), the data fit both uniform and non-uniform models similarly meaning that the available data appears inconclusive.

**Figure 1| The cumulative plot is insensitive to detect general heterogeneity in mutation rates**
*Figure demonstrating that, using a cumulative plot mutations, simulated by both highly heterogeneous mutation rates (green) and completely uniform mutation rates (purple) fit the uniform model when sorting by an uncorrelated heterogeneous model (dashed line). This reveals that the analysis by Maddamsetti et al. cannot detect whether mutation rates are uniform or heterogeneous in repair-deficient strains. At most, their result only reveals that the variation (if any) in repair-deficient strains does not correlate strongly with the variation that we report in wild-type strains, which is unsurprising given that both conditions are dominated by very different mutational processes.*

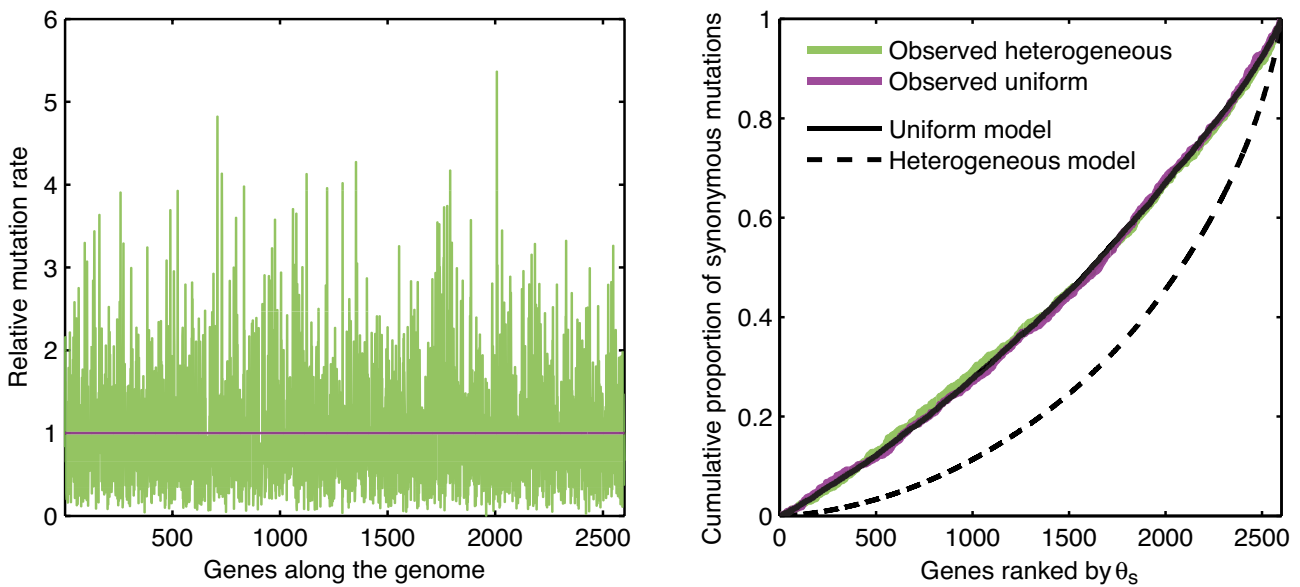

## 4. Preliminary evidence of non-randomness in the mutation-accumulation data

Maddamsetti's dataset of 1,069 synonymous mutations is also too small to perform any clean analysis at gene level, compared with our published dataset of >120,000 mutations. Since most genes have either zero (70.5%) or one (22.7%) mutations, any per-gene analysis will be extremely noisy.

Nevertheless, we performed some preliminary analyses that show tantalising evidence for non-randomness in Maddamsetti's dataset. There is an (unsurprisingly) noisy yet significant positive correlation between Maddamsetti's and our datasets ($P$=6.47e-05); this suggests that the underlying mutation rates are to some extent related even with the caveats above. Moreover, we also observe a significant negative correlation between the number of mutations in Maddamsetti's dataset and gene expression levels ($P$=3.15e-04), as we reported for the wildtype strains. This indicates that, despite the noisy nature of the underlying dataset, and the fact that repair-deficient conditions are not directly comparable to wildtype conditions, they show trends that are consistent with our own findings.

**Figure 2| Evidence of non-randomness of the mutation rate in repair-deficient strains**
*Left. Bar plot showing the median expression level of genes without a mutation (~70% of genes in Lenski's dataset) and the median expression level of genes with one or more mutations. Right. Bar plot showing the median expression level of the bottom (~70%) and top (~30%) mutator genes using θs' from Martincorena et al. as a reference to evaluate the left plot. Given that top mutators in the left are in most cases classified on the basis of a single mutation at best the result is expected to be extremely noisy. Thus we find this evidence very supportive of the association of the mutation rate with expression in these repair-deficient strains.*

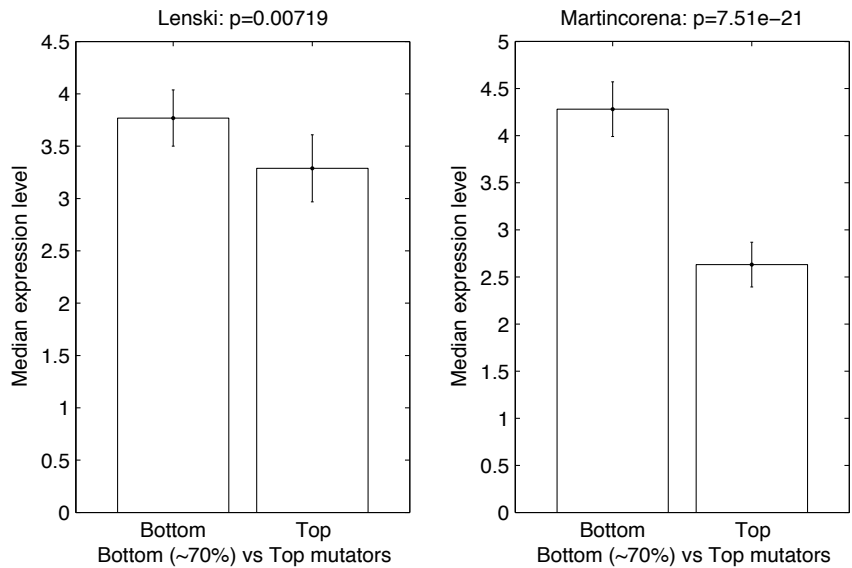

## 5. Conclusions

Having carefully evaluated the arguments and data from Maddamsetti *et al*. we conclude that:
- Mutations from repair-deficient strains cannot prove or disprove our results about the mutational processes dominating the spectrum of mutations in wild-type conditions.
- There is no evidence of HGT affecting our published estimates.
- Further analysis and most likely additional data are required to conclude whether mutation rates are uniform or not in repair-deficient strains.
- We find surprising preliminary evidence for non-randomness of mutations in repair-deficient strains.

If anything, these observations support our original conclusions. We hope that our comments and preliminary analyses help Maddamsetti et al. in making the most of their data.